\documentclass{article}

\usepackage{henry}

\title{Machine learning emulation of a local-scale UK climate model}

\author{%
  Henry Addison \\
  Department of Computer Science \\
  University of Bristol\\
  Bristol, UK \\
  \texttt{henry.addison@bristol.ac.uk} \\
  \And
  Elizabeth Kendon \\
  Met Office Hadley Centre \\
  Exeter, UK \\
  \And
  Suman Ravuri \\
  DeepMind \\
  London, UK \\
  \And
  Laurence Aitchison* \\
  Department of Computer Science \\
  University of Bristol \\
  Bristol, UK \\
  \And
  Peter AG Watson* \\
  School of Geographical Sciences \\
  University of Bristol \\
  Bristol, UK \\
 }

\begin{document}

\maketitle

\listoftodos
* equal contribution
\begin{abstract}
   Climate change is causing the intensification of rainfall extremes. Precipitation projections with high spatial resolution are important for society to prepare for these changes, e.g. to model flooding impacts. Physics-based simulations for creating such projections are very computationally expensive. This work demonstrates the effectiveness of diffusion models, a form of deep generative models, for generating much more cheaply realistic high resolution rainfall samples for the UK conditioned on data from a low resolution simulation. We show for the first time a machine learning model that is able to produce realistic samples of high-resolution rainfall based on a physical model that resolves atmospheric convection, a key process behind extreme rainfall. By adding self-learnt, location-specific information to low resolution relative vorticity, quantiles and time-mean of the samples match well their counterparts from the high-resolution simulation.
\end{abstract}

\section{Introduction}

Climate change is predicted to cause intensification of heavy rainfall extremes in the UK \cite{donat2016precipextreme, kendon2019ukcpscience}. Understanding rainfall at a local (\(\sim\) 2km) scale is important for better adapting to these changes - for example, to predict where flooding might occur, \cite{scahller2020resolutionrole}. Machine learning (ML) techniques can emulate high-resolution simulations using low-resolution inputs to generate local projections \cite{vandal2018mldownscaling, doury2022RCMemulator}. This means expensive projections from a climate model could be complemented with further, cheaper samples with realistic spatial and temporal structure enabling better understanding of the uncertainty of high-resolution precipitation. This is particularly useful for extreme events which are poorly sampled by high-resolution simulations due to their infrequent nature.

Numerical simulations of physical processes of the climate can be used to create projections of rainfall but they are extremely expensive leading to trade-offs. At one end of the scale are global climate models (GCMs). These produce global-covering projections. Many variants exist with large ensemble sizes which allow experimenting with different scenarios and uncertainty in the underlying physics and model. The trade-off is these models restrict their resolution to grids of 25km or more, which is too coarse to provide actionable insight \cite{gutierrez2019sdcomparison}.

At the other end of the simulation scale are dynamical downscaling models like the Met Office used to produce 2.2km local projections in their UKCP18 dataset \cite{ukcp18local}. These used a convection-permitting model (CPM), a specialised type of regional climate model (RCM), which had a fine enough resolution to produce local-scale projections that are useful to decision makers. The trade-off is a small ensemble size (12, all following a single climate change scenario following a single climate change scenario, SSP5-8.5) with reduced spatial and temporal domain (UK and Ireland for 60 years) \cite{kendon2019ukcpscience}.

An alternative approach to produce high resolution projections is called statistical downscaling. This approach attempts to fit a statistical relationship between a lower-resolution set of climate variables and higher-resolution values. They are much less computationally expensive than the dynamical downscaling approach of RCMs.

Many approaches have been tried based on both traditional statistics and machine learning. No method has proven better than the others in general \cite{gutierrez2019sdcomparison, vandal2018mldownscaling, fowler2007downscaling4hydrology}. \textcite{gutierrez2019sdcomparison} assess the skill of 45 different methods for precipitation alone for a single European collaborative experiment, though the better-performing methods relevant to this project's problem are mainly based on generalized linear models (GLMs) or nearest neighbours (methods of analogs). These only covered point-statistics at individual observation stations and cannot produce the full spatial structure of a rain field.

\textcite{doury2022RCMemulator} use U-Net \cite{Ronneberger2015unet} as the basis for their emulation of a 12km RCM based on GCM inputs over parts of France and Spain. \textcite{vandal2018mldownscaling} found no clear best option when comparing traditional approaches with some off-the-shelf machine learning approaches (Multi-task Sparse Structure Learning and Autoencoder Neural Networks) for daily and extreme precipitation over the Northeastern United States. These systems are deterministic: once trained, the same conditioning GCM input will produce the high-resolution output. Without a probabilistic element, these models struggle to predict the small-scale detail of precipitation \cite{ravuri2021deepgenprecip}. This work uses diffusion models \cite{song2019smld, ho2020ddpm, song2021sbgmsde} (aka score-based generative models), a recent development in deep generative models, to solve these problems with statistical downscaling.

Statistical downscaling of the full grid (rather than just a selection of observation stations) can be considered as similar to the super-resolution problem of natural images. The grid squares of the GCM and CPM outputs are like pixels with their intensity corresponding to climate variables such as daily rainfall (albeit not just integer values between 0 and 255). Common approaches for state-of-the-art sample generation, probability density estimation and super-resolution of natural images rely on deep generative models: diffusion models, Generative Adversarial Networks (GANs) \cite{goodfellow2014gans}, Variational Autoencoders (VAEs) \cite{kingma2014vaeorigin}, and autoregressive (AR) models like PixelRNN \cite{vandenoord2016pixelrnn}. Diffusion models have shown competitive results in the natural image domain \cite{dharwial2021diffbeatsgans} and offer desirable trade-offs in sample sharpness (VAEs tend to produce blurry samples) and sample diversity (GANs can suffer from mode collapse \cite{grover2018flowgan}) and sampling cost (computationally cheaper than AR and much cheaper than a simulation). For more details on diffusions models see Appendix \ref{app:diffmodels}.

Deep generative models have been successfully applied to problems in weather and climate such as short-term forecasting (nowcasting) of sequences of rainfall radar fields \cite{ravuri2021deepgenprecip} and downscaling rainfall \cite{Leinonen2020GANsd}. Commonly these use GANs but they can be difficult to train and their limitations imply they might underestimate the probability of extreme events. Diffusion models instead model the full data distribution and should not suffer these same problems. To our knowledge this work is the first application of diffusion models in the domain of climate downscaling.

\section{Method}

\textbf{Dataset} The Met Office's UKCP18 dataset contains both the low-resolution GCM global 60km projections \cite{ukcp18global} and high-resolution CPM local 2.2km projections \cite{ukcp18local}. Coarse relative vorticity at 850hPa is chosen as the conditioning input since it is a good predictor of high resolution rainfall \cite{chan2018precippredictors}. For practical purposes the target data cover a 64x64 region of an 8.8km grid (2.2km grid coarsened 4x). This is small enough to fit on available GPU, fine enough to be useful and covers a large enough area for samples to be interpretable. For more details of the data see Appendix \ref{app:datasets}.

\textbf{Model} The emulator is a score-based generative model based on NCSN++\cite{song2021sbgmsde}, adapted to allow conditional training and sampling. For sampling the Euler-Maruyama method is used to solve the reverse SDE. Location-specific parameters were added in order to tie pixels to the underlying physical location of each pixel. This allows the model to learn relevant features for each location which effect rainfall that may not be available from the coarse relative vorticity inputs alone. Two models were trained: one based on just the simulation data as the conditioning inputs and one with an 8-channel feature map of location-specific parameters that are learnt from the data stacked on top of the vorticity conditioning tensor. Implementation is available on Github: \url{https://github.com/henryaddison/score_sde_pytorch}, which is a fork of \url{https://github.com/yang-song/score_sde_pytorch} \cite{song2021sbgmsde}.

\section{Results} \label{sec:results}

\begin{figure}[h]
    \centering
    \begin{subfigure}{0.8\textwidth}
        \includegraphics[width=1.0\linewidth]{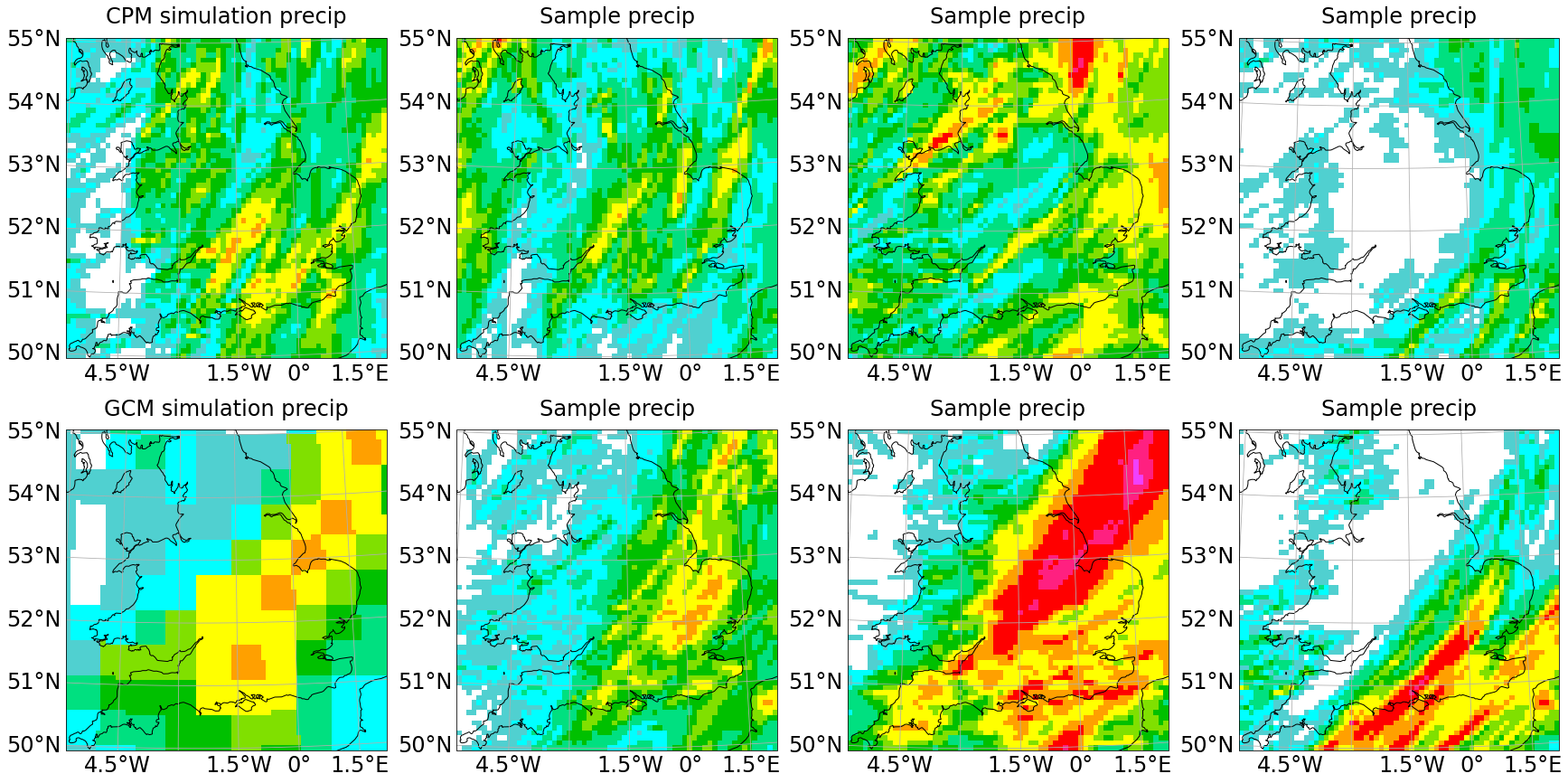}
    \end{subfigure}

    \begin{subfigure}{1.0\textwidth}
        \includegraphics[width=1.0\linewidth]{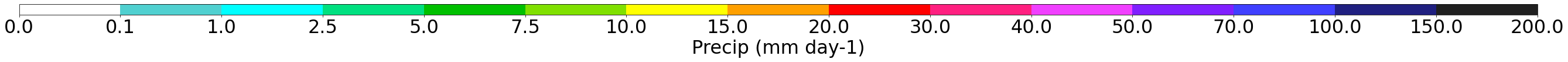}
    \end{subfigure}

    \caption{Example samples conditioned on data from CPM (top) and GCM (bottom). First column is the rainfall from the simulation of the conditioning input. Columns 2 to 4 are the samples from the ML model. The highly variable nature of rainfall means samples should not match the simulation rainfall but represent the full range of plausible rainfall for the given low resolution conditions, of which the simulation output is a single example.}
    \label{fig:results:samples}
\end{figure}

Figure \ref{fig:results:samples} shows example high resolution samples from a model conditioned on relative vorticity in both modes of operation: sampling based on conditioning input of coarsened CPM data or GCM data. For each model, 3 samples were generated for each of the 4,320 timestamps in the validation set. Note that the aim is for the model to capture the full range of rainfall that is possible given the low resolution conditions described by the relative vorticity. Therefore samples should look realistic and cover a range that plausibly includes the rainfall from the simulation but they shouldn't all match it since the simulation output is just one possible occurrence of a highly variable phenomenon.

\begin{figure}[h]
    \centering
    \includegraphics[width=0.7\linewidth]{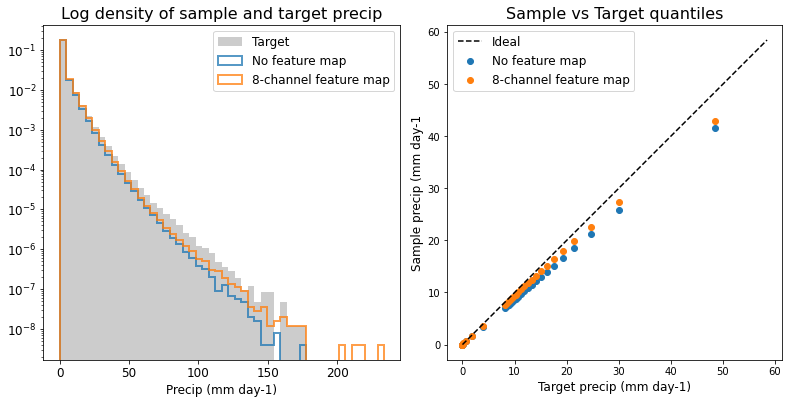}
    \caption{Per-pixel CPM-driven sample distribution: Log density (left) and Q-Q plot (right). Quantiles plotted are in three 9-member ranges of centiles: 10th-90th, 91st-99th, 99.1th-99.9th}
    \label{fig:results:cpm-pixeldist}
\end{figure}

Figure \ref{fig:results:cpm-pixeldist} shows the per-pixel Q-Q and log density plots of the sample and target distributions. These are based on samples conditioned on coarsened-CPM inputs for the two models variants. Both plots show the ML models match the target distribution well. The sample quantiles all match the target quantiles well and to be out by a constant scale factor of about 10\%. Results using GCM inputs are similar in quality (Appendix \ref{app:gcm-results}).

\begin{figure}[h]
    \begin{subfigure}{0.49\textwidth}
        \includegraphics[width=0.9\linewidth]{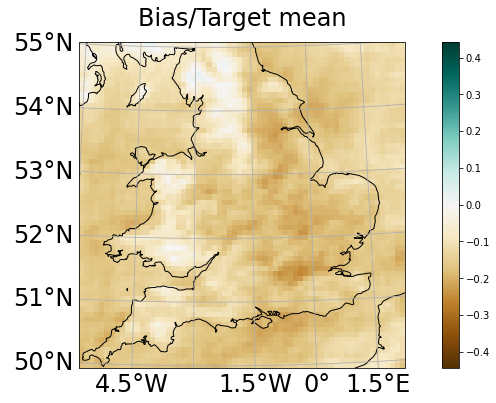}
        \caption{Coarsened-CPM-driven sample bias\\without location-specific parameters}
        \label{fig:results:cpm-bias:nofm}
    \end{subfigure}
    \begin{subfigure}{0.49\textwidth}
        \includegraphics[width=0.9\linewidth]{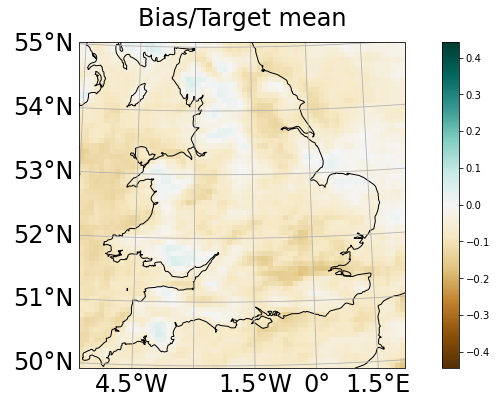}
        \caption{Coarsened-CPM-driven sample bias\\with location-specific parameters}
        \label{fig:results:cpm-bias:fm8}
    \end{subfigure}

    \caption{Mean-normalized bias using coarsened-CPM-based inputs with and without the self-learnt feature map}
    \label{fig:results:cpm-bias}
\end{figure}

Figure \ref{fig:results:cpm-bias} shows the normalized bias at each pixel. This is defined as sample mean (over time and all samples) minus target mean (over time) divided by the target mean (over time). The two versions represent without location-specific parameters (\ref{fig:results:cpm-bias:nofm}) and with self-learnt location-specific parameters (\ref{fig:results:cpm-bias:fm8}). The biases at each pixel are mostly negative so that on average the ML model is slightly under-predicting rainfall. Comparison between the two models show a reduction to the bias by adding the location-specific parameters. This matches the improvement in quantile results with the feature map (Figure \ref{fig:results:cpm-pixeldist}) and again the bias of this version lies within about 10\% of the target mean at each pixel.

\section{Conclusion} \label{sec:conclusion}

High-resolution projections of precipitation are needed to more effectively adapt to future changes in precipitation. However, to make these solely with numerical climate simulations is very expensive. A diffusion model designed for natural images has been adapted to create the first emulator of CPM rainfall. As a generative model, it can cope with the stochastic nature of rainfall and also produce more samples without needing new inputs. The result is cheaper, high resolution rainfall samples with realistic spatial structure based on low resolution relative vorticity.

Further improvements are possible too. Adding conditioning inputs based on temperature and humidity should provide more information than relative vorticity at 850hPa alone. The extrapolation properties of the ML model have not been tested to see how it copes with unseen, very extreme conditions well beyond the 99.9th centile. The performance of the model when transferred to generate samples based on GCM inputs could be improved by transforming the GCM variables to better match the distribution of the CPM variables. Further generalization to ensembles of other climate models will also be valuable, leading to the potential to use the ML model to downscale climate models for which physics-based downscaling has not been implemented. There may also be value in generating sequences rather than single snapshots of rainfall and higher temporal frequencies than daily.

\newpage
\printbibliography

@article{kendon2019ukcpscience,
   author = {Kendon, E J and Fosser, Giorgia and Murphy, James and Chan, Steven and Clark, Robin and Harris, Glen and Lock, Adrian and Lowe, Jason and Martin, Gill and Pirret, Jenny and others},
   title = {UKCP Convection-permitting model projections: Science report},
   year = {2019},
   url = {https://www.metoffice.gov.uk/pub/data/weather/uk/ukcp18/science-reports/UKCP-Convection-permitting-model-projections-report.pdf},
   type = {Journal Article}
}

@article{gutierrez2019sdcomparison,
   author = {Gutiérrez, J. M. and Maraun, D. and Widmann, M. and Huth, R. and Hertig, E. and Benestad, R. and Roessler, O. and Wibig, J. and Wilcke, R. and Kotlarski, S. and San Martín, D. and Herrera, S. and Bedia, J. and Casanueva, A. and Manzanas, R. and Iturbide, M. and Vrac, M. and Dubrovsky, M. and Ribalaygua, J. and Pórtoles, J. and Räty, O. and Räisänen, J. and Hingray, B. and Raynaud, D. and Casado, M. J. and Ramos, P. and Zerenner, T. and Turco, M. and Bosshard, T. and Štěpánek, P. and Bartholy, J. and Pongracz, R. and Keller, D. E. and Fischer, A. M. and Cardoso, R. M. and Soares, P. M. M. and Czernecki, B. and Pagé, C.},
   title = {An intercomparison of a large ensemble of statistical downscaling methods over Europe: Results from the VALUE perfect predictor cross-validation experiment},
   journal = {International Journal of Climatology},
   volume = {39},
   number = {9},
   pages = {3750-3785},
   DOI = {10.1002/joc.5462},
   year = {2019},
   type = {Journal Article}
}

@article{vandal2018mldownscaling,
   author = {Vandal, Thomas and Kodra, Evan and Ganguly, Auroop R.},
   title = {Intercomparison of machine learning methods for statistical downscaling: the case of daily and extreme precipitation},
   journal = {Theoretical and Applied Climatology},
   volume = {137},
   number = {1-2},
   pages = {557-570},
   DOI = {10.1007/s00704-018-2613-3},
   year = {2018},
   type = {Journal Article}
}

@article{fowler2007downscaling4hydrology,
   author = {Fowler, H. J. and Blenkinsop, S. and Tebaldi, C.},
   title = {Linking climate change modelling to impacts studies: recent advances in downscaling techniques for hydrological modelling},
   journal = {International Journal of Climatology},
   volume = {27},
   number = {12},
   pages = {1547-1578},
   DOI = {10.1002/joc.1556},
   year = {2007},
   type = {Journal Article}
}

@article{donat2016precipextreme,
   author = {Donat, Markus G. and Lowry, Andrew L. and Alexander, Lisa V. and O'Gorman, Paul A. and Maher, Nicola},
   title = {More extreme precipitation in the world's dry and wet regions},
   journal = {Nature Climate Change},
   volume = {6},
   pages = {508-513},
   DOI = {10.1038/nclimate2941},
   year = {2016},
   type = {Journal Article}
}

@article{chan2018precippredictors,
   author = {Chan, S. C. and Kendon, E. J. and Roberts, N. and Blenkinsop, S. and Fowler, H. J.},
   title = {Large-Scale Predictors for Extreme Hourly Precipitation Events in Convection-Permitting Climate Simulations},
   journal = {Journal of Climate},
   volume = {31},
   number = {6},
   pages = {2115-2131},
   ISSN = {0894-8755},
   DOI = {10.1175/Jcli-D-17-0404.1},
   year = {2018},
   type = {Journal Article}
}

@dataset{ukcp18local,
    author = {{Met Office Hadley Centre}},
    year = {2019},
    publisher = {Centre for Environmental Data Analysis},
    title = {{UKCP18 Local Projections at 2.2km Resolution for 1980-2080}},
    url = {https://catalogue.ceda.ac.uk/uuid/d5822183143c4011a2bb304ee7c0baf7},
    urldate = {2021-06-09}
}

@dataset{ukcp18global,
    author = {{Met Office Hadley Centre}},
    year = {2018},
    publisher = {Centre for Environmental Data Analysis},
    title = {{UKCP18 Global Projections at 60km Resolution for 1900-2100}},
    url = {https://catalogue.ceda.ac.uk/uuid/97bc0c622a24489aa105f5b8a8efa3f0},
    urldate = {2021-06-09}
}

@article{Leinonen2020GANsd,
   author = {Leinonen, Jussi and Nerini, Daniele and Berne, Alexis},
   title = {Stochastic Super-Resolution for Downscaling Time-Evolving Atmospheric Fields With a Generative Adversarial Network},
   journal = {IEEE Transactions on Geoscience and Remote Sensing},
   pages = {1–13},
   year = {2020},
   type = {Journal Article}
}

@article{grover2018flowgan,
   author = {Grover, A. and Dhar, M. and Ermon, S.},
   title = {Flow-GAN: Combining Maximum Likelihood and Adversarial Learning in Generative Models},
   journal = {Thirty-Second Aaai Conference on Artificial Intelligence / Thirtieth Innovative Applications of Artificial Intelligence Conference / Eighth Aaai Symposium on Educational Advances in Artificial Intelligence},
   volume = {32},
   number = {1},
   pages = {3069-3076},
   year = {2018},
   type = {Journal Article}
}

@inproceedings{vandenoord2016pixelrnn,
   author = {Van Den Oord, Aäron and Kalchbrenner, Nal and Kavukcuoglu, Koray},
   title = {Pixel recurrent neural networks},
   booktitle = {33rd International Conference on Machine Learning, ICML 2016},
   volume = {4},
   pages = {2611-2620},
   year = {2016},
   type = {Conference Proceedings}
}

@article{ravuri2021deepgenprecip,
   author = {Ravuri, Suman and Lenc, Karel and Willson, Matthew and Kangin, Dmitry and Lam, Remi and Mirowski, Piotr and Fitzsimons, Megan and Athanassiadou, Maria and Kashem, Sheleem and Madge, Sam and Prudden, Rachel and Mandhane, Amol and Clark, Aidan and Brock, Andrew and Simonyan, Karen and Hadsell, Raia and Robinson, Niall and Clancy, Ellen and Arribas, Alberto and Mohamed, Shakir},
   title = {Skillful Precipitation Nowcasting using Deep Generative Models of Radar},
   DOI = {arxiv:2104.00954},
   year = {2021},
   type = {Journal Article}
}

@article{Ronneberger2015unet,
   author = {Ronneberger, O. and Fischer, P. and Brox, T.},
   title = {U-Net: Convolutional Networks for Biomedical Image Segmentation},
   journal = {Medical Image Computing and Computer-Assisted Intervention, Pt Iii},
   volume = {9351},
   pages = {234-241},
   ISSN = {0302-9743},
   DOI = {10.1007/978-3-319-24574-4_28},
   year = {2015},
   type = {Journal Article}
}

@article{kingma2014vaeorigin,
   author = {Kingma, Diederik P and Welling, Max},
   title = {Auto-Encoding Variational Bayes},
   year = {2014},
   type = {Journal Article}
}

@inproceedings{song2021sbgmsde,
   author = {Song, Yang and Sohl-Dickstein, Jascha and Diederik and Kumar, Abhishek and Ermon, Stefano and Poole, Ben},
   title = {Score-Based Generative Modeling through Stochastic Differential Equations},
   booktitle = {ICLR},
   DOI = {arxiv:2011.13456},
   type = {Conference Proceedings},
   year = {2021}
}

@inproceedings{dharwial2021diffbeatsgans,
   author = {Dhariwal, Prafulla and Nichol, Alexander},
   title = {Diffusion models beat GANs on image synthesis},
   booktitle = {Advances in Neural Information Processing Systems},
   volume = {34},
   url = {https://proceedings.neurips.cc/paper/2021/file/49ad23d1ec9fa4bd8d77d02681df5cfa-Paper.pdf},
   year = {2021},
   type = {Conference Proceedings}
}

@inproceedings{ho2020ddpm,
   author = {Ho, Jonathan and Jain, Ajay and Abbeel, Pieter},
   title = {Denoising diffusion probabilistic models},
   booktitle = {Advances in Neural Information Processing Systems},
   volume = {33},
   pages = {6840-6851},
   url = {https://proceedings.neurips.cc/paper/2020/hash/4c5bcfec8584af0d967f1ab10179ca4b-Abstract.html},
   type = {Conference Proceedings}
}

@inproceedings{song2019smld,
   author = {Song, Yang and Ermon, Stefano},
   title = {Generative modeling by estimating gradients of the data distribution},
   booktitle = {Advances in Neural Information Processing Systems},
   volume = {32},
   url = {https://proceedings.neurips.cc/paper/2019/hash/3001ef257407d5a371a96dcd947c7d93-Abstract.html},
   type = {Conference Proceedings}
}

@article{doury2022RCMemulator,
   author = {Doury, Antoine and Somot, Samuel and Gadat, Sebastien and Ribes, Aurélien and Corre, Lola},
   title = {Regional climate model emulator based on deep learning: concept and first evaluation of a novel hybrid downscaling approach},
   journal = {Climate Dynamics},
   ISSN = {0930-7575
1432-0894},
   DOI = {10.1007/s00382-022-06343-9},
   url = {https://doi.org/10.1007/s00382-022-06343-9},
   year = {2022},
   type = {Journal Article}
}

@inproceedings{goodfellow2014gans,
   author = {Goodfellow, Ian J. and Pouget-Abadie, Jean and Mirza, Mehdi and Xu, Bing and Warde-Farley, David and Ozair, Sherjil and Courville, Aaron and Bengio, Yoshua},
   title = {Generative adversarial nets},
   booktitle = {Proceedings of the 27th International Conference on Neural Information Processing Systems - Volume 2},
   publisher = {MIT Press},
   pages = {2672–2680},
   year = {2014},
   type = {Conference Proceedings}
}

@article{scahller2020resolutionrole,
   author = {Schaller, N. and Sillmann, J. and Muller, M. and Haarsma, R. and Hazeleger, W. and Hegdahl, T. J. and Kelder, T. and van den Oord, G. and Weerts, A. and Whan, K.},
   title = {The role of spatial and temporal model resolution in a flood event storyline approach in western Norway},
   journal = {Weather and Climate Extremes},
   volume = {29},
   pages = {100259},
   ISSN = {2212-0947},
   DOI = {ARTN 100259
10.1016/j.wace.2020.100259},
   year = {2020},
   type = {Journal Article}
}

\newpage
\appendix
\section{Diffusion Models} \label{app:diffmodels}

Probabilistic models assume that observed data, such as high-resolution rainfall over the UK, is drawn from an unknown distribution \(p^*(\mathbf{x})\). A conditional model such as high-resolution rainfall conditioned on coarse GCM inputs \(p^*(\mathbf{x}|\mathbf{g})\) can also be considered but for simplicity this section will stick with the context free version.

\textcite{song2021sbgmsde} combine earlier approaches \cite{song2019smld, ho2020ddpm} into a single framework called Score-Based Generative Models with Stochastic Differential Equations (SDEs). The idea is to imagine a diffusion process \(\{\mathbf{x}(t)_{t=0}^{T}\}\) modelled by an SDE:

\begin{equation}
  \mathrm{d}\mathbf{x} = \mathbf{f}(\mathbf{x}, t)\mathrm{d}t + g(t)d\mathbf{w}
\end{equation}

When run forward a sample, \(\mathbf{x}(0)\), from the data distribution, \(p_0\), is gradually perturbed over time into a sample from a final noise distribution, \(p_T\), somewhat like how a structured gas sample will gradually diffuse randomly across a room. The final distribution is chosen as something tractable for sampling, usually a Gaussian.

More interesting for us is running the diffusion process backwards:

\begin{equation} \label{eqn:mlbg:reversesde}
  \mathrm{d}\mathbf{x} = [\mathbf{f}(\mathbf{x}, t) - g(t)^2 \nabla_{\mathbf{x}}\log{p_{t}(x)}]{d}t + g(t)d\bar{\mathbf{w}}
\end{equation}

By solving this, samples from \(p_T\) (which is easy by design) can be converted into samples from the original data distribution. This requires two steps: calculating the score, \(\nabla_{\mathbf{x}}\log{p_{t}(x)}\), and then applying numerical approaches to solve Equation \ref{eqn:mlbg:reversesde}.

The score is estimated as a neural net \(s_\theta(\mathbf{x}, t)\) where \(\theta\) are determined by minimizing:

\begin{equation}
  \mathbb{E}_t \{
    \lambda(t) \mathbb{E}_{\mathbf{x}(0)} \mathbb{E}_{\mathbf{x}(0) | \mathbf{x}(t)}
      \left[
        || s_\theta(\mathbf{x}(t), t) -  \nabla_{\mathbf{x}(t)}\log{p_{0t}(\mathbf{x}(t) | \mathbf{x}(0))} ||_2^2
      \right
    \}
\end{equation}
where \(\lambda\) is a positive weighting function that is chosen along with f and g.

\textcite{song2021sbgmsde} summarize three approaches for solving the reverse SDE. General-purpose numerical methods can be used to find approximate solutions to the SDE. Predictor-Corrector sampling takes this a step further by using making use of estimated score at each timestep to apply a correction to the sample estimated at that timestep by the general purpose solver. Alternatively the problem can be reformulated as a deterministic process without affecting the trajectory probabilities and in turn solved using an ODE solver.

\section{Dataset} \label{app:datasets}

The Met Office's UKCP18 datasets for both the low-resolution GCM global 60km projections \cite{ukcp18global} and high-resolution CPM local 2.2km projections \cite{ukcp18local}. For the whole UK and Ireland domain, the high-resolution grid is 484 x 606 while the coarse grid is 17 x 23. For practical purposes this study limits the geographical domain to the area covered by a 64x64 area of an 8.8km grid (so 2.2km high-resolution coarsened 4x) centred on Birmingham. This covers a large enough area (England and Wales) to make sample interpretation easier while also boosting the resolution of the 60km projections. The temporal frequency is daily. Data are available hourly for three 20-year periods (1981-2000, 2021-2040, 2061-2080). The default CPM ensemble member and its corresponding GCM were chosen. The dataset is split into a training set (70\%, n=15,120) and a validation set (20\%, n=4,320) with a 10\% test set reserved once fine-tuning is complete.

It is not possible to train a model using GCM data as conditioning input because the weather conditions in GCM and CPM are different for the same timestamp \cite{doury2022RCMemulator}. As far are the CPM is concerned, the GCM stops at the edge of Europe so the two simulations diverge at times. Therefore, the dataset used in the training phase will the conditioning input data is from the CPM regridded and coarsened to match the output of the GCM and the target is 4x coarsened precipitation from the CPM. It is also better preferable to avoid using low resolution rainfall as a conditioning input but rather variables that are as well-represented in GCM as the CPM, such as wind, temperature and humidity. Based on work by \textcite{chan2018precippredictors}, the first conditioning input variable was chosen to be relative vorticity (a derivative of wind field which measures circulation in the atmosphere and correlated with how stormy or fair conditions are) at altitude corresponding to 850hPa. Once fitted, samples can then be drawn from the model either using conditioning inputs based on coarsened CPM data or data from a GCM.

\section{GCM-driven results} \label{app:gcm-results}

\begin{figure}[h]
    \centering
    \includegraphics[width=0.7\linewidth]{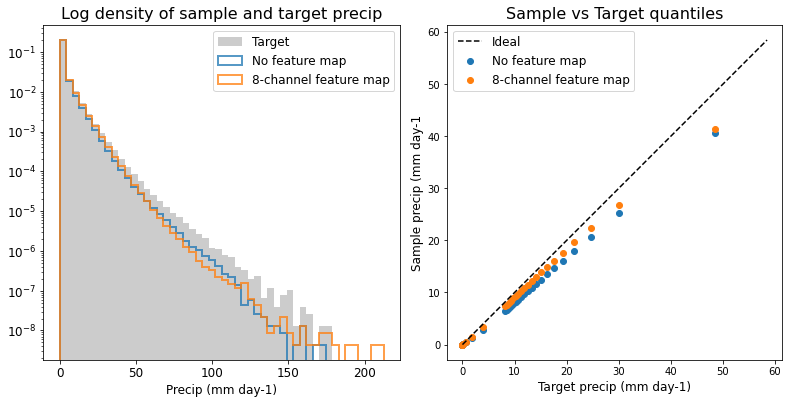}
    \caption{Per-pixel GCM-driven sample distribution: Log density (left) and Q-Q plot (right)}
    \label{fig:results:gcm-pixeldist}
\end{figure}

\begin{figure}[h]
    \begin{subfigure}{0.49\textwidth}
        \includegraphics[width=0.9\linewidth]{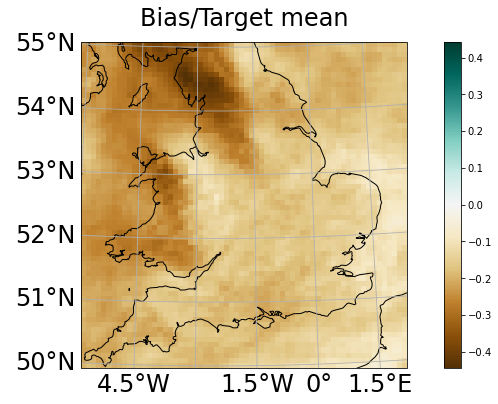}
        \caption{GCM-driven sample bias without self-learnt\\feature map}
        \label{fig:results:gcm-bias:nofm}
    \end{subfigure}
    \begin{subfigure}{0.49\textwidth}
        \includegraphics[width=0.9\linewidth]{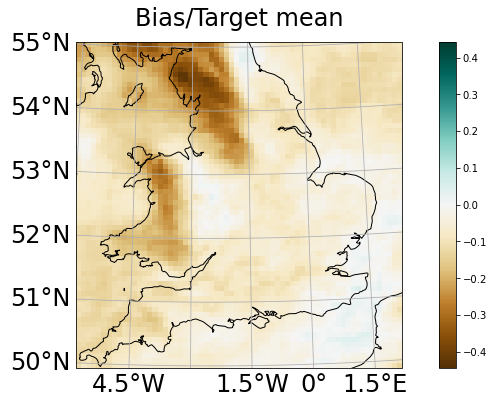}
        \caption{GCM-driven sample bias with an 8-channel\\self-learnt feature map}
        \label{fig:results:gcm-bias:fm8}
    \end{subfigure}

    \caption{Mean-normalized bias using GCM inputs with and without the self-learnt feature map}
    \label{fig:results:gcm-bias}
\end{figure}

Comparing the bias between sampling modes (CPM-based samples in \ref{fig:results:cpm-bias:fm8} with GCM-based samples in \ref{fig:results:gcm-bias:fm8})  shows a difference between the sample distributions depending on the source of the conditioning input. The CPM-based bias is smaller overall but there is also an inversion: CPM-based bias is smallest (and slightly positive) over upland areas like Snowdonia and the Lake District but GCM-based bias is much larger (and negative) other these same upland region.

\end{document}